\documentclass[aps,prx,reprint,notitlepage,nofootinbib,eqsecnum,twocolumn]{revtex4-2}
\usepackage[utf8]{inputenc}

\usepackage{physics}
\usepackage{amstext}
\usepackage{amsfonts,amssymb,mathtools,amsmath,graphicx}
\usepackage{bbm}

\usepackage{indentfirst}

\usepackage{mathrsfs}
\usepackage[normalem]{ulem}
\usepackage{color}
\usepackage{amsthm}
\usepackage{dcolumn}% Align table columns on decimal point
\usepackage{bm}% bold math
\usepackage[utf8]{inputenc}
\usepackage[T1]{fontenc}
\usepackage{booktabs, array, 
lipsum, multirow}
\usepackage[dvipsnames]{xcolor}
\usepackage{tikz}
\usepackage[bottom, perpage]{footmisc}

\usepackage{hyperref}
\usepackage{orcidlink}
\usepackage{cleveref}

\newcommand{\be}{\begin{equation}}
\newcommand{\ee}{\end{equation}}
\crefname{figure}{Fig.}{Fig.}
\crefname{equation}{Eq.}{Eq. }
\crefname{section}{Sec.}{Secs.}

\usetikzlibrary{decorations.markings, arrows.meta}
\tikzset{
    mag/.style={
        draw=Maroon,thick,
        postaction={decorate, decoration={
            markings,
            mark=at position 0.6 with {\arrow{Stealth}}
        }}
    }
}
\tikzset{
    mab/.style={
        draw=MidnightBlue,thick,
        postaction={decorate, decoration={
            markings,
            mark=at position 0.6 with {\arrow{Stealth}}
        }}
    }
}
\usetikzlibrary{decorations.pathmorphing}
\usetikzlibrary{backgrounds}

\begin{document}

\title{The holographic dual of the GHZ state}

\author{Libo Jiang\orcidlink{0009-0008-7942-1874}}
\email{sustech@buaa.edu.cn}
\author{Yan Liu}
\email{yanliu@buaa.edu.cn}

\affiliation{
\vspace{0.1cm}
\mbox{Center for Gravitational Physics, Department of Space Science}\\ 
\mbox{and International Research Institute
of Multidisciplinary Science, }\\
\mbox{Beihang University, Beijing 100191, China}
\\
\mbox{Peng Huanwu Collaborative Center for Research and Education,} \\
\mbox{Beihang University, Beijing 100191, China}}

\begin{abstract}
Current holographic research mostly focuses on a subset of quantum systems with a classical gravity dual. Although the precise boundary of this subset remains unknown, certain constraints are recognized; for instance, holographic entropies obey inequalities that are violated by general quantum states such as the GHZ states. This paper, however, proposes a gravity dual for the GHZ states—a non-manifold geometry termed the \textit{booklet wormhole}. We demonstrate an exact match for all entropy properties with the GHZ state, as well as the identity of the Euclidean partition functions for both systems. The booklet wormhole circumvents the conventional holographic entropy inequalities because different topologies are inevitably included in the gravitational path integral, even in the large-$N$ limit. This provides the first explicit holographic duality with a non-perturbative quantum effect. Remarkably, the construction is simple, and the dual state is maximally entangled, making it ideal for gedanken experiments.
\end{abstract}

\maketitle
\section{Introduction}
Holography establishes a duality between quantum systems and higher-dimensional quantum gravity systems, with AdS/CFT being the most well-known example. Currently, most bottom-up research on holography is restricted to ``holographic'' quantum systems from the outset, as not all quantum systems or states have a classical gravity dual. For example, the holographic entropy (also known as the Ryu-Takayanagi entropy \cite{06RTEntro,06RTEntR,09RTEntR}) imposes constraints on holographic states. Specifically, Ref.~\cite{13HolIne} proved that tripartite information is non-positive for all holographic states. Yet, this inequality fails to hold for certain quantum states, notably the GHZ states \cite{89GHZ,90GHZ}. Therefore, it was believed that these GHZ states do not have a holographic dual, at least not a classical dual \cite{25NoGHZ}.  
Nevertheless, we propose that GHZ states have a holographic dual: a non-manifold geometry with non-trivial topology, which we term the \textit{booklet wormhole}. The holographic entropies of this geometry precisely match those of the GHZ states, and both systems share identical Euclidean partition functions.

Previously, holography research was largely confined to a few geometries: vacuum AdS \cite{98AdSCFT}, two-sided black holes \cite{03TFDTSB}, and their modifications. The GHZ state/booklet wormhole duality establishes a new testing ground for holography. Beyond being the first known duality that violates the conventional holographic entropy bound, the GHZ state also serves as a representative state for many-body entanglement.  This duality will therefore play a fundamental role in the quantum information perspective of holography. Research on two-sided black holes could now be revisited in the context of the booklet wormhole, advancing our understanding of many-body correlation beyond pairwise entanglement.

\section{Preliminary}
\subsection{GHZ state}
Greenberger, Horne, and Zeilinger first constructed the 4-spin GHZ state in \cite{89GHZ} as a generalization of the Bell state. In the following year, \cite{90GHZ} considered the 3-qubit GHZ state, which could be the most common form of the GHZ state:
\begin{equation}
|\text{GHZ}\rangle=\frac{1}{\sqrt{2}}(\ket{000}+\ket{111}).
\end{equation} 
The 3-qubit GHZ state holds particular significance because \cite{00GHZ-W} demonstrated that three qubits admit only two fundamentally distinct entanglement classes under LOCC (local operations and classical communication) equivalence. The GHZ state serves as the maximally entangled prototype for one class, while the W state $\ket{W} = \frac{1}{\sqrt{3}} (\ket{001} + \ket{010} + \ket{100})$ represents the other class.

It can be generalized to the $n$-partite, $D$-dimensional GHZ state:
\begin{equation}
|\text{GHZ}_n\rangle=\frac{1}{\sqrt D}\sum_{i=0}^{D-1} (\ket{i}^{\bigotimes n} ).
\end{equation}
Every $D$-dimensional subsystem is maximally mixed with entropy $\ln D$. Furthermore, any combination of $m$ subsystems shares the same entropy for $1\leq m\leq n-1$.

For the purpose of this work, we introduce the \textit{thermal GHZ state}:
\begin{equation}
\label{eq:tghz}
\ket{\beta\text{-GHZ}_n} \coloneqq\frac{1}{\sqrt Z}
\sum_{i=0}^{D-1} \left( e^{-\frac{\beta E_i}{2}}\ket{i}^{\bigotimes n} \right),
\end{equation}
where $\ket{i}$ are the energy eigenstates with energies $E_i$, $\beta$ is a real parameter that is referred to as the inverse temperature, and $Z$ is a normalization factor. The reduced density matrix for each subsystem is 
\begin{equation}
\rho=\frac{1}{Z}\sum_{i=0}^{D-1}  e^{-\beta E_i} \ket{i}\bra{i},
\end{equation} which justifies the interpretation of $\beta$. The original GHZ states can be regarded as the infinite-temperature limit of the thermal GHZ states.

All states possessing a classical gravity dual must obey the monogamy condition of mutual information: their tripartite information is non-positive \cite{13HolIne,12HolIne}, 
\be \begin{aligned}
I_3(A:B:C) \coloneqq& S(A)+S(B)+S(C)-S(AB)\\
&-S(BC)-S(AC)+S(ABC)\leq 0.\label{ineq}\end{aligned}
\ee 
Nevertheless, both GHZ states and thermal GHZ states can violate this inequality.
For the $n$-partite GHZ/thermal GHZ state, all nontrivial subsystems share the same entanglement entropy, so we can find subsystems $A$, $B$, $C$ such that $I_3(A:B:C)=S(A)=S(B)=S(C)>0$ for the case $n\geq 4$.

\subsection{Booklet geometry}
The booklet geometry generalizes the gluing of spacetime boundaries.
When we want to glue two spacetime boundaries together, the variation of their total gravitational action typically does not vanish at the interface. We need a brane with the appropriate energy-momentum to satisfy the on-shell condition $\delta I_\text{total}=0$. This variational principle yields the \textit{Israel junction conditions} \cite{66JunCon}:
\begin{gather} h^{(1)}_{\mu\nu}=h^{(2)}_{\mu\nu},\\
 \Delta K_{\mu\nu}-\Delta Kh_{\mu\nu}=-8\pi GT_{\mu\nu},
\end{gather}
where $h_{\mu\nu}^{(i)}$ is the induced metric of the $i$-th boundary, $\Delta K_{\mu\nu}$ denotes the extrinsic curvature difference between two boundaries with its trace $\Delta K$, and $T_{\mu\nu}$ is the brane's energy-momentum tensor. 

Ref.~\cite{24BookL,24BookD} generalized the original bipartite Israel junction condition to the $n$-partite case, where $n$ spacetime boundaries are glued along a common interface. The resulting geometry is termed a ``\textit{booklet}''. A brane similarly resides on the common interface, and the generalized junction condition can be derived from varying the Einstein-Hilbert action containing the Gibbons-Hawking-York boundary terms:
\begin{gather}
h^{(1)}_{\mu\nu}=h^{(2)}_{\mu\nu}=
\cdots=h^{(n)}_{\mu\nu}, \label{I1}\\
\sum_i K^{(i)}_{\mu\nu}-K^{(i)}h_{\mu\nu}=-8\pi GT_{\mu\nu},\label{I2}
\end{gather}
where $h^{(i)}_{\mu\nu}$ and $K^{(i)}_{\mu\nu}$ are the induced metric and the extrinsic curvature of the $i$-th ``page'', respectively, and $T_{\mu\nu}$ is the energy-momentum tensor of the brane. This junction condition inherently exhibits $S_n$ symmetry. Note that a geometry with such a multi-way junction cannot constitute a differentiable manifold, as the dimension of the tangent space jumps discontinuously at the junction. However, the GHZ state is known to lack a manifold geometry dual, whereas the holographic dictionary suggests a holographic dual for all quantum states. This motivates us to explore the holographic duality beyond classical geometries. Other applications of booklet geometry in holography can be found in \cite{25refl1,26holnet}.

\subsection{Multi-entropy}
Multi-entropy \cite{22NEntro} generalizes the conventional bipartite entanglement entropy to quantify multi-partite entanglement.
The conventional entanglement entropy can be calculated from a single subsystem's density matrix \cite{09RTEntR}:
\be
S\coloneqq\lim_{\alpha\rightarrow 1}\frac{1}{1-\alpha} \ln \frac{Z_\alpha}{Z_1^\alpha}, 
\ee where $Z_\alpha=\text{Tr}  \tilde \rho^\alpha$ contains $\alpha$ copies of the unnormalized density matrix $\tilde \rho$ (we use a tilde to label unnormalized operators or states). 

When we have an $n$-partite system, the entanglement can be fully described by the density matrix of $n-1$ subsystems, which contains $2(n-1)$ indices, 
\be\begin{aligned}
&\tilde \rho_{aa'bb'\cdots mm'}\\
&\coloneqq\bra{\phi_{a_1}}\bra{\phi_{b_2}}\cdots \bra{\phi_{m_{n-1}}} \hat {\tilde \rho}\ket{\phi_{{a'}_1}}\ket{\phi_{{b'}_2}}\cdots \ket{\phi_{{m'}_{n-1}}} ,
\end{aligned}\ee
where $\ket{\phi_{i_j}}$ represents the $i$-th basis vector of the $j$-th subsystem. 
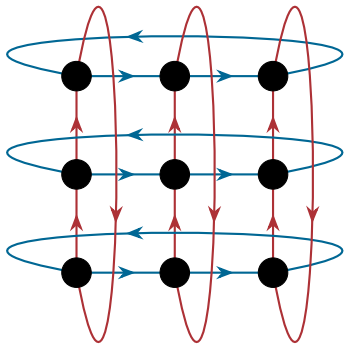
\begin{figure}
    \centering
    \begin{tikzpicture} [scale=1.3]            
    \useasboundingbox (-1,-1) rectangle (3,3);
        
        \draw [mag] (0,0)--(0,1);\draw [mag] (0,1)--(0,2);
        \draw [mag] (1,0)--(1,1);\draw [mag] (2,1)--(2,2);
        \draw [mag] (2,0)--(2,1);\draw [mag] (1,1)--(1,2);
        \draw [mab] (0,0)--(1,0);\draw [mab] (1,0)--(2,0);
        \draw [mab] (0,1)--(1,1);\draw [mab] (1,2)--(2,2);
        \draw [mab] (0,2)--(1,2);\draw [mab] (1,1)--(2,1);
        \draw [mab] (2,0) to [out=10, in=170, looseness=4] (0,0);
        \draw [mab] (2,1) to [out=10, in=170, looseness=4] (0,1);
        \draw [mab] (2,2) to [out=10, in=170, looseness=4] (0,2);   
        \draw [mag] (0,2) to [out=80, in=280, looseness=4] (0,0);
        \draw [mag] (1,2) to [out=80, in=280, looseness=4] (1,0);
        \draw [mag] (2,2) to [out=80, in=280, looseness=4] (2,0);   
        \filldraw (0,0) circle (.15);
        \filldraw (0,1) circle (.15);
        \filldraw (0,2) circle (.15);
        \filldraw (1,0) circle (.15);
        \filldraw (1,1) circle (.15);
        \filldraw (1,2) circle (.15);
        \filldraw (2,0) circle (.15);
        \filldraw (2,1) circle (.15);
        \filldraw (2,2) circle (.15);
        %\pgfresetboundingbox
    \end{tikzpicture}
    \caption{The figure illustrates the tensor network that calculates $Z_{3^2}$. Each dot represents a copy of $\rho_{ii'jj'}$, with four outer lines corresponding to four indices, and different colors represent different subsystems. The arrows point from a bra to a ket with the same index; summing over all possible indices gives $Z_{3^2}$.}
    \label{fignentr}
\end{figure}
The $\alpha$ copies of a two-index density matrix can build a one-dimensional tensor network giving $Z_\alpha$. Similarly, $\alpha^{n-1}$ copies of a $(2n-2)$-index density matrix can build an $(n-1)$-dimensional (hyper)cubic tensor network giving $Z_{\alpha^{n-1}}$, illustrated in \cref{fignentr}. 
This $Z_{\alpha^{n-1}}$ generates the \textit{multi-entropy}:\footnote{Ref.~\cite{22NEntro} defines the Rényi multi-entropy to be $\frac{1}{1-\alpha} \ln \frac{Z_{\alpha^{n-1}}}{Z_1^{\alpha^{n-1}}}$ rather than using a denominator proportional to ${1-\alpha^{n-1}}$. This could cause the multi-entropy to diverge in the large-$\alpha$ limit.}
\begin{equation}
S^{(n)}\coloneqq\lim_{\alpha\rightarrow 1} 
\frac{1}{1-\alpha}  
\ln  \frac{Z_{\alpha^{n-1}}}{Z_1^{\alpha^{n-1}}}.
\end{equation}

The multi-entropy for thermal GHZ states \eqref{eq:tghz} has a simple expression \cite{22NEntro}, 
\begin{equation}
\label{eq:nentropy-tghz}
S^{(n)}=(n-1)S^{(2)},
\end{equation} where $S^{(n)}$ denotes the $n$-entropy under arbitrary $n$-partitioning of the thermal GHZ state, and $S^{(2)}$ represents the von Neumann entropy of any single subsystem. 

Ref.~\cite{22NEntro} constructed an example demonstrating that multi-entropy can distinguish states with identical bipartite entanglement entropies. Consider 
the \textit{W state} \be \frac{1}{\sqrt 3} (\ket{001}+\ket{010}+\ket{100})\ee and a thermal GHZ state \be \frac{1}{\sqrt 3}(\sqrt{2} \ket{000}+\ket{111}). \ee For arbitrary bipartitions, the reduced density matrices of both states share identical spectra, implying that bipartite entanglement entropies fail to distinguish them. However, the W and GHZ states have different entanglement structures, revealed through differences in their 3-entropies. This exemplifies that multi-entropies are essential for identifying GHZ-type entanglement.\footnote{The inequality for $I_3$ or any bipartite entanglement entropy is insufficient to exclude the tripartite GHZ state. However, a recent study \cite{25NoGHZ} introduced an inequality incorporating multi-entropy and demonstrated that the tripartite GHZ state also cannot have a classical holographic dual.}

In holography, the $n$-entropy admits a simple geometric dual analogous to the RT surface: it equals $\frac{1}{4G}$ times the minimal area of the bulk surfaces that $n$-partition the boundary \cite{22NEntro}. We refer to these partitioned bulk regions as the \textit{multipartite entanglement wedge} of the corresponding boundary system.\footnote{The physical interpretation of the multi-partite entanglement wedge remains poorly understood.}

\section{Geometry dual to the thermal GHZ state}

Every single subsystem of the thermal GHZ state has a reduced density matrix identical to that of the thermofield double state at the same temperature. This implies that both states are indistinguishable under observations limited to a single subsystem. Consequently, the gravity dual of their subsystems should share identical entanglement wedges according to entanglement wedge reconstruction \cite{15BulRec}. For a sufficiently high temperature admitting a stable black hole, geometry is precisely the AdS-Schwarzschild geometry outside the horizons,
\be\label{eq:metric0}
ds^2=-f(r)dt^2+\frac{dr^2}{f(r)}+r^2d\Omega_{d-1}^2 ,
\ee
\be 
f(r)
=\begin{cases} 
   \dfrac{r^2}{l^2}+1-\dfrac{16\pi GM}{(d-1)V_\Omega r^{d-2}}, & \text{if } d>2, \\[8pt]
   \dfrac{r^2}{l^2}-8\pi GM, & \text{if } d=2 ,
\end{cases}\label{ffunc}
\ee where $V_\Omega$ is the volume of the $d-1$ dimensional unit sphere, $l$ is the AdS radius, and $M$ is the ADM mass. 

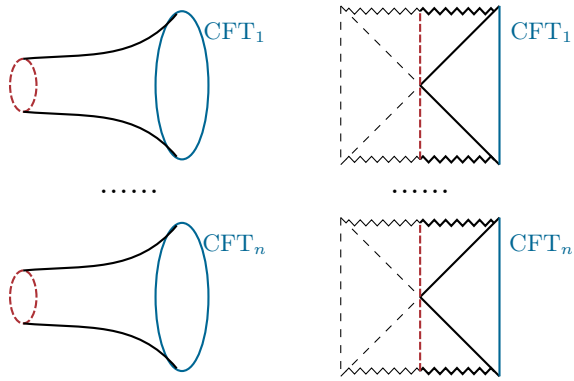
\begin{figure}
\centering
\begin{tikzpicture}[scale=0.7]
\draw[dash pattern=on 3pt off 1pt,color=Maroon,thick] (-3cm,0) ellipse (0.25cm and 0.5cm); 
\draw[color=MidnightBlue,thick](0,0) ellipse (0.5cm and 1.4cm); 
\draw[thick] (-3,0.5) to[out=5,in=230] (-0.1cm,1.35cm);
\draw[thick] (-3,-0.5) to[out=355,in=130] (-0.1cm,-1.35cm);
\node[color=MidnightBlue] at (1,1) {CFT$_1$};
\draw[dashed] [-] (3,1.5)--(3,-1.5);
\draw[decorate, decoration={zigzag, segment length=1.5mm, amplitude=0.4mm}] (3.1,1.4) -- (4.5,1.4);
\draw[decorate,thick, decoration={zigzag, segment length=1.5mm, amplitude=0.4mm}] (4.5,1.4) -- (5.9,1.4);
\draw[decorate, decoration={zigzag, segment length=1.5mm, amplitude=0.4mm}] (3.1,-1.4) -- (4.5,-1.4);
\draw[decorate,thick, decoration={zigzag, segment length=1.5mm, amplitude=0.4mm}] (4.5,-1.4) -- (5.9,-1.4);
\draw[dashed,color=black] [-] (3cm,1.5cm)--(4.5cm,0cm);
\draw[color=black,thick] [-] (4.5cm,0cm)--(6cm,-1.5cm);
\draw[dashed,color=black] [-] (3cm,-1.5cm)--(4.5cm,0cm);
\draw[color=black,thick] [-] (4.5cm,0cm)--(6cm,1.5cm);
\draw[dash pattern=on 3pt off 1pt,color=Maroon,thick] [-] (4.5cm,-1.4cm)--(4.5cm,1.4cm);
\node[color=MidnightBlue] at (6.8cm,1.0cm) {CFT$_1$};
\draw[color=MidnightBlue,thick] [-] (6cm,-1.5cm)--(6cm,1.5cm);

\begin{scope}[shift={(0,3)}]
\node at(-1,-5) {\Large......};
\draw[dash pattern=on 3pt off 1pt,color=Maroon,thick] (-3cm,-7) ellipse (0.25cm and 0.5cm); 
\draw [color=MidnightBlue,thick](0,-7) ellipse (0.5cm and 1.4cm); 
\draw[thick] (-3,-6.5) to[out=5,in=230] (-0.1cm,-5.65cm);
\draw[thick] (-3,-7.5) to[out=355,in=130] (-0.1cm,-8.35cm);
\node[color=MidnightBlue] at (1,-6) {CFT$_n$};

\node at(4.5,-5) {\Large......};
\draw[dashed] [-] (3,-5.5)--(3,-8.5);
\draw[decorate, decoration={zigzag, segment length=1.5mm, amplitude=0.4mm}] (3.1,-5.6) -- (4.5,-5.6);
\draw[decorate,thick, decoration={zigzag, segment length=1.5mm, amplitude=0.4mm}] (4.5,-5.6) -- (5.9,-5.6);
\draw[decorate, decoration={zigzag, segment length=1.5mm, amplitude=0.4mm}] (3.1,-8.4) -- (4.5,-8.4);
\draw[decorate,thick, decoration={zigzag, segment length=1.5mm, amplitude=0.4mm}] (4.5,-8.4) -- (5.9,-8.4);
\draw[dashed,color=black] [-] (3cm,-5.5cm)--(4.5cm,-7cm);
\draw[color=black,thick] [-] (4.5cm,-7cm)--(6cm,-8.5cm);
\draw[dashed,color=black] [-] (3cm,-8.5cm)--(4.5cm,-7cm);
\draw[color=black,thick] [-] (4.5cm,-7cm)--(6cm,-5.5cm);
\draw[dash pattern=on 3pt off 1pt,color=Maroon,thick] [-] (4.5cm,-8.4cm)--(4.5cm,-5.6cm);
\node[color=MidnightBlue] at (6.8cm,-6cm) {CFT$_n$};
\draw[color=MidnightBlue,thick] [-] (6cm,-8.5cm)--(6cm,-5.5cm);
\end{scope}
\end{tikzpicture}
\caption{The illustrations of a booklet wormhole. \textit{Left:} the $t=0$ time slice of the 2+1-dimensional booklet wormhole. Red dashed circles indicate horizon cuts, where each segment corresponds to one half of a two-sided black hole. These segments are joined through a multipartite junction condition. \textit{Right:} The Penrose diagram of the booklet wormhole. Red dashed lines denote cuts that pass bifurcation surfaces of Killing horizons. There is a family of reflection symmetry planes related by boost symmetry, and we may select an arbitrary one for the cut.
}
 \label{fig}
\end{figure}

All the differences arise behind the horizons. Since the GHZ states have $S_n$ permutation symmetry, the bulk dual must exhibit the same symmetry. We propose a geometry by preparing $n$ copies of the geometry and connecting them symmetrically by a multi-way junction:
\begin{enumerate}
    \item Cut each two-sided black hole along a timelike reflection-symmetry plane, retaining half the spacetime;
    \item Glue $n$ such copies along their cuts using the booklet geometry, with the junction satisfying the $n$-partite junction condition \cref{I1,I2}.
\end{enumerate} 
We refer to the resulting geometry as a \textit{booklet wormhole}, illustrated in \cref{fig}.
The junction condition \cref{I1} is automatically satisfied since we select $n$ identical copies. While we could have considered a general cut satisfying \cref{I2}, we instead chose the reflection symmetry plane as the cut. This reflection symmetry guarantees vanishing extrinsic curvature at all boundaries, implying $T_{\mu\nu}=0$ through \cref{I2}. We will demonstrate in the next section why the GHZ state requires a junction free of energy and momentum.

\section{Euclidean path integral}
The Euclidean path integral that prepares the thermofield double state defines the boundary conditions for the Euclidean black hole, and their partition functions are identical, $Z_\text{CFT}=Z_\text{AdS}$. This supports the statement that the two-sided black holes are dual to the thermofield double states. We will prove that the Euclidean preparation of the thermal GHZ state similarly matches the Euclidean continuation of the booklet wormhole.

\subsection{CFT side}
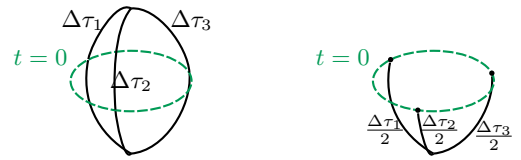
\begin{figure}
    \centering
    \begin{tikzpicture}[scale=0.8]
        \node at (-0.8,1) {$\Delta \tau_1$};
        \node at (-0,0) {$\Delta \tau_2$};
        \node at (1,1) {$\Delta \tau_3$};
        \draw[color=ForestGreen,thick,dash pattern=on 4pt off 1pt] (0,0) ellipse (1 and 0.5);
        \node[color=ForestGreen,thick] at (-1.5,0.4) {$t=0$};
        \node[circle,fill=black,inner sep=0pt,minimum size=0pt] (a) at (0.966,0.13) { };
        \node[circle,fill=black,inner sep=0pt,minimum size=0pt] (b) at (-0.708,0.354){ };
        \node[circle,fill=black,inner sep=0pt,minimum size=0pt] (c) at (-0.258,-0.482){ };

        \draw[thick, black] (0,1.2) to[out=5, in=94,looseness=0.78]  (a);
        \draw[thick, black] (0,1.2) to[out=-210, in=79, looseness=0.66]  (b);
        \draw[thick, black] (0,1.2) to[out=-130, in=94, looseness=0.7]  (c);
        \begin{scope}[on background layer]
        \draw[thick, black] (0,-1.2) to[out=5, in=-86,looseness=0.87]  (a);
        \draw[thick, black] (0,-1.2) to[out=-210, in=-101, looseness=0.99]  (b);
        \draw[thick, black] (0,-1.2) to[out=-130, in=-86, looseness=0.6]  (c);
        \end{scope}
        
 \begin{scope}[xshift=2cm]
                \draw[color=ForestGreen,thick,dash pattern=on 4pt off 1pt] (3,0) ellipse (1 and 0.5);
        \node[color=ForestGreen,thick] at (1.5,0.4) { $t=0$};
        \node[circle,fill=black,inner sep=0pt,minimum size=2pt] (a) at (3.966,0.13) { };
        \node[circle,fill=black,inner sep=0pt,minimum size=2pt] (b) at (-0.708+3,0.354){ };
        \node[circle,fill=black,inner sep=0pt,minimum size=2pt] (c) at (-0.258+3,-0.482){ };
        \node at (2.2,-0.8) {$\frac{\Delta \tau_1}{2}$};
        \node at (3.1,-0.8) {$\frac{\Delta \tau_2}{2}$};
        \node at (4,-0.9) {$\frac{\Delta \tau_3}{2}$};
        \begin{scope}[on background layer]
        \draw[thick, black] (3,-1.2) to[out=5, in=-86,looseness=0.87]  (a);
        \draw[thick, black] (3,-1.2) to[out=-210, in=-101, looseness=0.99]  (b);
        \draw[thick, black] (3,-1.2) to[out=-130, in=-86, looseness=0.6]  (c);
        \end{scope}
        \end{scope}
    \end{tikzpicture}

    \caption{The figures illustrate a Euclidean preparation of the $3$-partite thermal GHZ state, displaying only the imaginary time dimension. The left panel depicts the geometric representation of the partition function, where three black lines of length $\Delta\tau_i$ correspond to three periods of Euclidean time evolution. Cutting along the time-reflection symmetry plane ($t=0$) yields a CFT state, as shown in the right panel.}
    \label{figeucft}
\end{figure}

Consider a geometry comprising $n$ Euclidean CFTs, each with its two ends respectively connecting to two junctions (see \cref{figeucft}). We define the direction between two ends as the Euclidean time direction. Each CFT possesses a distinct temporal interval $\Delta\tau_i$ between the two ends. Each junction can be dealt with as a path integral region with $n$ open boundaries; we propose that the corresponding states of both junctions are the (unnormalized) GHZ state, $\ket{\widetilde{\text{GHZ}} } =\sum_i\ket{i}^{\bigotimes n}.$ 
The reflection symmetry of the geometry enables us to perform a cut along the symmetry plane ($t=0$ slice), where each half prepares an (unnormalized) state,\footnote{Our construction implicitly singles out a preferred basis. A detailed discussion about this can be found in our subsequent paper \cite{26NoLoJu}.}  
\be
\ket{\tilde \phi}=\sum_i e^{-\frac{\Delta\tau_1}{2} E_i}e^{-\frac{\Delta\tau_2}{2} E_i}...e^{-\frac{\Delta\tau_n}{2} E_i} \ket{i}^{\bigotimes n},
\ee
at the $t=0$ surface.
Different path integrals can prepare the same thermal GHZ state $\ket{\widetilde {\beta\text{-GHZ}_n}}
=\sum_{i}  e^{-\frac{\beta}{2}E_i}\ket{i}^{\bigotimes n}$ as long as $\sum_i\Delta\tau_i={\beta}$. 

The partition function of the thermal GHZ state is given by 
\be\label{ZCFT}
Z_{\beta\text{-GHZ}_n}\coloneqq \bra{\widetilde {\beta\text{-GHZ}}_n}\ket{\widetilde {\beta\text{-GHZ}}_n}=\text{tr}\left(e^{\sum_i \Delta\tau_i H}\right),
\ee which is identical to that of a thermofield double state at the same temperature.

This partition function implies that the \textit{boundary entropy} $\ln g$ \cite{91BouEnt,04BCFT} of the GHZ-state junctions vanishes:
\be \ln g=\frac{1}{2} 
\lim_{\substack{\Delta\tau_i \to \infty \\ i = 1,\dots,n}}\ln \frac{Z_{\beta\text{-GHZ}_n}}{\prod_i Z_{\Delta\tau_i}}=0,\ee where $Z_{\Delta\tau_i}$ denotes the partition function for a defect-free CFT with imaginary time period $\Delta\tau_i.$ We will not delve into the details of boundary entropy. For our purposes, it suffices to note that when $\ln g = 0$, the AdS interfaces extending from the boundary junctions have vanishing extrinsic curvature and are vertical to the boundary \cite{08BouEnt,11BCFT2}.\footnote{If the junction can deform freely with no energy cost, then its exact shape is physically irrelevant. Nevertheless, for convenience in our derivation, we may still treat this case as vertical to the boundary.} This vertical relation is essential for our construction in the next subsection.

\subsection{AdS side}
The booklet wormhole consists of $n$ identical pages, each described by the AdS-Schwarzschild metric. Consequently, the Euclidean continuation within each page corresponds to a section of a Euclidean black hole. 
The metric of the Euclidean AdS-Schwarzschild black hole is
\be
ds^2=f(r)d\tau^2+\frac{dr^2}{f(r)}+r^2d\Omega_{d-1}^2,
\ee
where $f(r)$ is defined in \cref{ffunc}. The imaginary time is periodic, $\tau\sim\tau+\beta$, where $\beta$ is the inverse temperature of the dual CFT. 

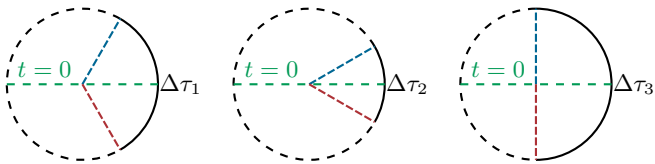
\begin{figure}
    \centering
    \begin{tikzpicture}
     \begin{scope}[xshift=-0cm]
     \draw [thick](0.5,0.866) arc (60:-60:1);
     \draw [thick,dashed](0.5,0.866) arc (60:300:1);
     \draw [thick,MidnightBlue,dash pattern=on 3pt off 1pt] (0,0)--(0.5,0.866);
     \draw [thick,Maroon,dash pattern=on 3pt off 1pt] (0,0)--(0.5,-0.866);
     \node at (1.3,0){$\Delta \tau_1$};
     \draw[dashed,thick,ForestGreen] (-1,0)--(1,0);
     \node[ForestGreen] at(-0.5,0.2){$t=0$};
     \end{scope}
        
     \draw [thick](3+0.866,0.5) arc (30:-30:1);
     \draw [thick,dashed] (3+0.866,0.5) arc (30:330:1);
     \draw [thick,MidnightBlue,dash pattern=on 3pt off 1pt] (3,0)--(3+0.866,0.5);
     \draw [thick,Maroon,dash pattern=on 3pt off 1pt] (3,0)--(3+0.866,-0.5);
     \node at (4.3,0){$\Delta \tau_2$};
     \draw[dashed,ForestGreen,thick] (2,0)--(4,0);
     \node[ForestGreen] at(2.5,0.2){$t=0$};

     \begin{scope}[xshift=0cm]
     \draw [thick](6,1) arc (90:-90:1);
     \draw [thick,dashed] (6,1) arc (90:270:1);
     \draw [thick,MidnightBlue,dash pattern=on 3pt off 1pt] (6,0)--(6,1);
     \draw [thick,Maroon,dash pattern=on 3pt off 1pt] (6,0)--(6,-1); 
     \node at (7.3,0){$\Delta \tau_3$};
     \draw[dashed,ForestGreen,thick] (5,0)--(7,0);
     \node[ForestGreen] at(5.5,0.2){$t=0$};
      \end{scope}
    \end{tikzpicture}
    \caption{The figure illustrates a possible Euclidean extension of a 3-page booklet wormhole. Each circle represents a Euclidean black hole. We cut along the constant-$\tau$ slices marked by red/blue dashed lines and glue the cuts with the same color.}
    \label{figeucads}
\end{figure}

As mentioned, the CFT junctions require the corresponding bulk interfaces to be vertical to the boundary, i.e., lying on constant-$\tau$ slices. To satisfy the boundary conditions from CFT, we excise each Euclidean black hole along two constant-$\tau$ slices separated by an interval $\Delta\tau_i$. We then collectively identify one constant-$\tau$ cut of each black hole, while identifying the other constant-$\tau$ cuts separately. This geometric construction is illustrated in \cref{figeucads}. 
All constant-$\tau$ hypersurfaces share the same induced metric and possess vanishing extrinsic curvature due to the $\tau$-translation and $\tau$-reflection symmetries of the metric. Through the generalized Israel conditions \cref{I1,I2}, these symmetries enable gluing along constant-$\tau$ slices with a vanishing stress-energy tensor at the junction.

 However, the presence of a kink at the Euclidean horizon—where the interfaces meet—complicates the analysis. The aforementioned symmetries do not guarantee that $T_{\mu\nu}=0$ at this juncture. Furthermore, as noted in Ref. \cite{94AddAct}, the total gravitational action cannot always be treated as a simple summation of the individual page actions, especially when boundaries exhibit discontinuous kinks. In such cases, an additional residual term must be subtracted upon identifying boundaries. A new prescription is required to evaluate these residual terms at multi-way junctions.

We aim to construct a bulk action consistent with the behavior of the boundary theory. To this end, consider a boundary partition function:
\be Z=\sum_{ij} \bra{i}_1\bra{i}_2\bra{i}_3 e^{-H_1\Delta \tau_1} e^{-H_2\Delta \tau_1} e^{-H_3\epsilon}  \ket{j}_1\ket{j}_2\ket{j}_3,
\ee where $\epsilon$ is a small parameter. In the limit $\epsilon\rightarrow 0$, $Z\rightarrow \sum_{ij} \bra{i}_1\bra{i}_2 e^{-H_1\Delta \tau_1} e^{-H_2\Delta \tau_1} \ket{j}_1\ket{j}_2$. Physically, this implies that appending a sector of vanishing opening angle does not alter the total action. Within this infinitesimal sector, both the Einstein-Hilbert and Gibbons-Hawking-York terms vanish, leaving only finite contributions from the three Hayward joint terms located at the corners of the sector.
The Hayward joint term \cite{93HayTer} is given by 
\be
I_{\text{Hayward}} = \frac{1}{8\pi G} \int_{\Sigma} \Theta \, d\Sigma
\ee where $d\Sigma$ is the proper area element on the codimension-2 joint, and $\Theta$ is the joint angle between the normals of the two intersecting hypersurfaces. Consequently, the total action of an infinitesimal sector is given by:
\be I_\text{infinitesimal sector}=I_\text{horizon}+2 I_\text{boundary}=  \pi\frac{A_h}{8\pi G}+  \pi\frac{A_b}{8\pi G},
\ee where $A_h$ is the area of the Euclidean horizon, and $A_b$ is the spatial area of the AdS boundary (evaluated at the radial cutoff). Since the insertion of this infinitesimal sector must leave the total action invariant, we must subtract a residual term when attaching a new sector via the multi-way junction:
\be I^\text{multi-way}_\text{residue}=I_\text{infinitesimal sector}=\pi\frac{A_h}{8\pi G}+  \pi\frac{A_b}{8\pi G},\ee
such that the total action after gluing is:
\be I_\text{total}=I_\text{original}+I_\text{sector}-I^\text{multi-way}_\text{residue}.\ee We emphasize that this construction is designed specifically for the joint of bulk interfaces dual to the Euclidean GHZ junctions, and may not apply to general multi-way junctions.

\begin{figure}
    \centering
    \begin{tikzpicture}[scale=1.8]
       \draw [thick](1,0) arc (0:-180:1); 
       \draw [thick,MidnightBlue,dash pattern=on 3pt off 1pt] (0,0)--(1,0);
       \draw [thick,Maroon,dash pattern=on 3pt off 1pt] (0,0)--(-1,0);
       \draw [thick](1.1,0.2) arc (0:60:1);
       \draw [thick,MidnightBlue,dash pattern=on 3pt off 1pt] (0.1,0.2)--(0.1+0.5,1.066);
       \draw [thick,Maroon,dash pattern=on 3pt off 1pt] (0.1,0.2)--(1.1,0.2);
       \draw [thick](-1.1,0.2) arc (180:60:1);
       \draw [thick,Maroon,dash pattern=on 3pt off 1pt] (-0.1,0.2)--(-0.1+0.5,1.066);
       \draw [thick,MidnightBlue,dash pattern=on 3pt off 1pt] (-0.1,0.2)--(-1.1,0.2);
       \node at (0,-1.16) {$\Delta\tau_3$};
       \node at (-0.75,1.2) {$\Delta\tau_1$};
       \node at (1.2,0.8) {$\Delta\tau_2$};
    \end{tikzpicture}
    \caption{Rearrange the pieces in the \cref{figeucads}, where each represents a sector of the Euclidean black hole geometry. When the circular sectors have appropriate radii, they can combine to form a perfect disk. The resulting geometry is a Euclidean black hole without defect. }
    \label{figtangram}
\end{figure}
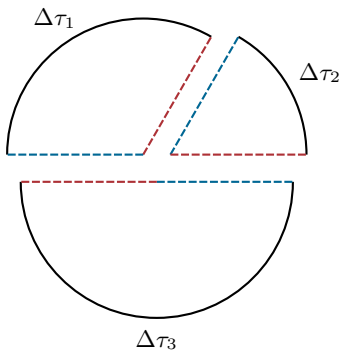

Next, we need to compute the action of each piece of the Euclidean booklet wormhole. A convenient approach is to use a ``tangram trick'' to reassemble these pieces into a well-studied geometry (\cref{figtangram}), adding the corresponding residual terms to obtain the total action of sectors.

The individual pieces can be reassembled into a complete, smooth Euclidean black hole because $\sum_i \Delta\tau_i=\beta$ (\cref{figtangram}). The residual term arising from gluing $n$ sectors ($n\geq3$) into a complete disk can be calculated using standard methods \cite{94AddAct}:
\be I_\text{residue}^\text{disk}(n)= (n-2)\pi\frac{A_h}{8\pi G}+  n\pi\frac{A_b}{8\pi G}.\ee
Thus, the total action of the sectors is related to the action of the Euclidean black hole by $I^\text{sectors}(n)=I^\text{blackhole}+I_\text{res}^\text{disk}(n).$

To construct a Euclidean booklet wormhole, we first glue two sectors into a disk (generically with a conical singularity) with a standard residual term, and then attach the remaining sectors via a multi-way junction. The total residual contribution is given by
\be \begin{aligned}I_\text{residue}^\text{booklet}(n)&=I_\text{residue}^\text{disk}(2)+(n-2)I^\text{multi-way}_\text{residue}\\&=(n-2)\pi\frac{A_h}{8\pi G}+  n\pi\frac{A_b}{8\pi G}.\end{aligned}\ee
Notably, the total residual term for gluing a Euclidean booklet wormhole is identical to that of a Euclidean black hole; consequently, the total action of the Euclidean booklet wormhole equals that of the Euclidean black hole at the same temperature. This is expected, since the Euclidean booklet wormhole is holographically dual to the thermal GHZ state, which shares the same partition function as a thermofield double state at the same temperature. Furthermore, the Hayward contributions localized at the Euclidean horizon and at the cutoff joints are exactly canceled by the residual terms. Consequently, within this multi-way gluing prescription, no additional codimension-two source is required at the Euclidean horizon.

\section{Holographic entropy of the booklet wormholes}

In this section, we compute the holographic entropy and multi-entropy of the booklet wormholes to verify their consistency with the GHZ states.
The covariant holographic entropy \cite{07HRT} requires finding the extremal surface with the minimal area. Utilizing the time-reflection symmetry, this reduces to finding minimal surfaces within the $t=0$ symmetry plane. 

Within the replica-trick framework, the RT surface arises as the interface separating bulk regions associated with distinct group elements \cite{13RepTri,22NEntro,23FunRep}. This point deserves some clarification. Consider a bipartition of the real line: the origin must be assigned exclusively to one of the two subsets. If, instead, the origin is treated as a separate point, the resulting partition $(-\infty,0)\cup {0} \cup(0,+\infty)$ is in fact a tripartition rather than a bipartition. The RT surface behaves analogously: it cannot sit precisely at the junction, but must instead be locally contained within a single page of the booklet geometry. Since each page is itself a manifold with boundary, the standard RT prescription applies directly within it. One should note, however, that the area of the homologous surfaces is discontinuous across the junction, since each horizon there is adjacent to $n-1$ other horizons. To pin down the location of the RT surface unambiguously, we define $\mathcal{S}^{(i)}$ as the junction interface lying on the $i$-th page.

\subsection{Three-page case}

\begin{figure}
\begin{tikzpicture}
   \centering
\draw[very thick,color=Maroon] [-] (0,0)--(0,1.7);
\node[color=Maroon] at (-0.3,1.4) {$A$};
\draw[very thick,color=black] (0,0)--(-1.4,-1.4);
\node[color=black] at (-1.3,-0.8) {$B$};
\draw[very thick,color=black]  (0,0)--(1.4,-1.4);
\node[color=black] at (1.3,-0.8) {$C$};
\draw [fill,color=black] (0,0) circle [radius=0.1];
\end{tikzpicture}
\caption{The figure illustrates the entanglement wedge of a single subsystem in the $t=0$ timeslice of a 3-page booklet wormhole. Only the radial direction is displayed, so each point represents a codimension-2 sphere. The central dot corresponds to the junction. The red line depicts the entanglement wedge of subsystem $A$, and the RT surface (the interface between this entanglement wedge and the complementary bulk region) contains a single external surface of the junction.}
    \label{fig3-2}
\end{figure}
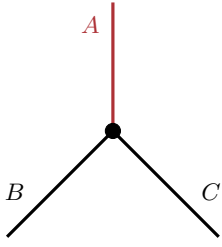
We begin by considering a 3-page booklet wormhole. \cref{fig3-2} shows that the minimal surface belonging to the $i$-th subsystem is the $\mathcal S^{(i)}$, whose area equals the horizon area $A_h$. Thus, the entanglement entropy of any subsystem is 
\be\label{eq:ee2-3page}
S^{(2)}=\frac{A_h}{4G} .\ee

 \begin{figure}
    \centering
    \begin{tikzpicture}
\draw[very thick,color=Maroon] [-] (6,0)--(6,1.7);
\node[color=Maroon] at (-0.3+6,1.4) {$A$};
\draw[very thick,color=black] (0+6,0)--(-1.4+6,-1.4);
\node[color=black] at (-1.3+6,-0.8) {$B$};
\draw[very thick,color=MidnightBlue]  (0+6,0)--(1.4+6,-1.4);
\node[color=MidnightBlue] at (1.3+6,-0.8) {$C$};
\draw [fill,color=Maroon] (0+6,0) circle [radius=0.1];
\end{tikzpicture}
    \caption{The figure illustrates a tripartition corresponding to the 3-entropy $S^{(3)}(A:B:C)$. Different multi-partite entanglement wedges are labeled by different colors. The interface of the tripartition consists of the two external surfaces of the junction.}
    \label{fig3-3}
\end{figure}
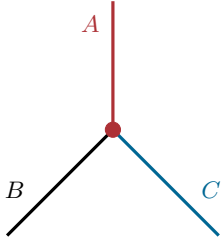
The 3-partite entropy is proportional to the area of the minimal tripartition surface \cite{22NEntro,23FunRep}. \Cref{fig3-3} shows the minimal surface for 3-partite entropy $S(A:B:C)$, where the junction may reside within the 3-partite entanglement wedge of any subsystem.
Non-unique choices of the minimal surface add an $O(1)$ term to the entropy, which is ignored in the large-$N$/small-$G$ limit. 
The minimal tripartition surface contains two $\mathcal S^{(i)}$, which yields 
\be S^{(3)}=2\frac{A_h}{4G}=2S^{(2)},
\label{eq:s3-hol}
\ee 
agreeing with the entropy structure of the 3-partite GHZ states in \cref{eq:nentropy-tghz}. 

\subsection{Case of four pages or more} \label{nonclassical}
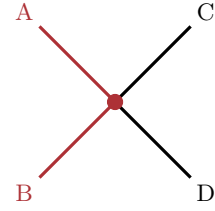
\begin{figure}
    \centering
    \begin{tikzpicture}
     
     \draw [Maroon,very thick] (-1,-1) to (0,0);
     \draw [Maroon,very thick] (-1,1) to (0,0);
     \draw [very thick] (1,1) to (0,0);
     \draw [very thick] (1,-1) to (0,0);
     \draw [fill,color=Maroon] (0,0) circle [radius=0.1];
     \node [Maroon] at(-1.2,-1.2) {B};
     \node [Maroon] at(-1.2,1.2) {A};
     \node  at(1.2,1.2) {C};
     \node  at(1.2,-1.2) {D};
    \end{tikzpicture}
    \caption{A plausible way to calculate the RT entropy of the subsystem $A\cup B$. The red region represents the entanglement wedge of subsystems $A\cup B$. The RT surface contains two external surfaces of the junction.}
    \label{fig4-2wrong}
\end{figure}

 We move to the 4-page case, where a subtlety arises. When calculating the holographic entropy for two subsystems, say $A$ and $B$, it seems that the RT surface contains two $\mathcal S^{(i)}$, as \cref{fig4-2wrong} suggests. This would yield an entropy of  $2\frac{A_h}{4G}=2S(A)=2S(B),$
 contradicting the GHZ states' expectation $S(A B)=S(A)=S(B)$. 
 
However, the above calculation is incorrect. Recall the replica trick, 
\be
S\coloneqq-\partial_\alpha \ln \text{tr}{\rho^\alpha}|_{\alpha=1}=-\partial_\alpha \ln \frac{Z_\alpha}{Z_1^\alpha}.
\ee
In the large-$N$/small-$G$ limit, we only consider the saddle point contribution of the bulk partition function, so that 
\begin{align} 
Z_1=e^{-I_1},\quad Z_\alpha=e^{-I_\alpha},
\end{align}where $I_1$ and $I_\alpha$ represent the Euclidean on-shell action of the single geometry and the $\alpha$-replica geometry, respectively.
An on-shell solution means $\delta I=0$ for a small variation of the metric. We will see that while the geometry in \cref{fig4-2wrong} satisfies the classical equation of motion and gives $I_1$, the corresponding replicated geometry does not constitute a valid saddle point for $I_\alpha$: an infinitesimal deformation of the geometry would significantly reduce its action.

We modify the original 4-page wormhole by replacing its codimension-1 junction with a finite-thickness region. Specifically, supposing the original junction has the metric $ds^2=g_{\mu\mu'}dx^\mu dx^{\mu'}$, we retain these metric terms while introducing an extra coordinate $\xi\in[0,1]$ parameterizing the thickness, where $\xi=0 \text{ and } \xi=1$ represent two external boundaries of the thickened junction. The additional metric components are set to $g_{\xi\mu}=0$ and $g_{\xi\xi}=\epsilon$, with $\epsilon$ a small parameter. The original codimension-1 junction is recovered when $g_{\xi\xi}=0$. 

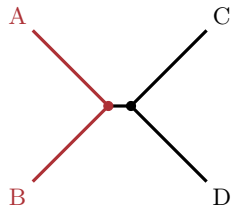
\begin{figure}
    \centering
    \begin{tikzpicture}
     
     \draw [Maroon,very thick] (-1,-1) to (0,0);
     \draw [Maroon,very thick] (-1,1) to (0,0);
     \draw [very thick] (0.3,0) to (0,0);
     \draw [very thick] (1.3,1) to (0.3,0);
     \draw [very thick] (1.3,-1) to (0.3,0);
     \draw [fill,color=Maroon] (0,0) circle [radius=0.06];
      \draw [fill] (0.3,0) circle [radius=0.06];
     \node [Maroon] at(-1.2,-1.2) {B};
     \node [Maroon] at(-1.2,1.2) {A};
     \node  at(1.5,1.2) {C};
     \node  at(1.5,-1.2) {D};
    \end{tikzpicture}
    \caption{The geometry that dominates the partition function $Z_\alpha$ in the calculation of $S(A B)$. The red lines represent a possible entanglement wedge of subsystems $A\cup B$.  The short line in the middle represents the junction with a small thickness.  The original $S_4$ symmetry is broken by the entropy considered.}
    \label{fig4-2right}
\end{figure}

 Pages connected to the junction now adhere to one of the two surfaces, $\xi=0 \text{ or } \xi=1$. Different gluing configurations yield different geometries, each with an individual minimal surface.
 \cref{fig4-2right} shows one such configuration that gives the minimal ``minimal surface''. The resulting entropy is $\frac{A_h}{4G}$, consistent with the expectation from the GHZ states. 

Why does this modified geometry yield the correct entropy, if it is indeed the right answer? The classical limit of $Z_\alpha$ requires minimizing  $I_\alpha$, where \be I_\alpha=\alpha I_1+ \frac{A}{4G} d\alpha+O(d\alpha^2)\ee for small $d\alpha\coloneqq\alpha-1$, with $A$ the area of a candidate RT surface. To compute the von Neumann entropy, we only focus on the limit $\alpha\rightarrow 1$, where $\alpha I_1$ dominates $I_\alpha$. Consequently, the standard procedure first minimizes $I_1$ and then minimizes $A$, corresponding to finding the minimal surface within the on-shell geometry. This process typically gives the correct minimum of $I_\alpha$.   

However, for booklet wormholes, introducing a small variation $\epsilon$ only incurs an infinitesimal increase in $I_1$, yet dramatically reduces $A$. Specifically, for arbitrarily small $d\alpha$, we can always find an $\epsilon$ such that the action $I_\alpha$ decreases:\be\Delta I_\alpha=\frac{\Delta A}{4G}d\alpha+\alpha\Delta I_1<0,\ee where $\Delta A=-\frac{A_h}{4G}$ quantifies the area difference between the minimal surfaces depicted in \cref{fig4-2right} and \cref{fig4-2wrong}, and $\Delta I_1>0$ is the action increment due to introducing a non-zero $\epsilon$. Therefore, it is worth modifying the geometry when calculating $Z_\alpha$. 

\begin{figure}
    \centering
    \begin{tikzpicture}
     \draw [Maroon,very thick] (-0.7,-1.5)node[below] {$A_1$} -- (0,0);
     \draw [Maroon,very thick] (0,-1.5) node[below] {$A_2$}-- (0,0);
     \draw [Maroon,very thick] (0.7,-1.5)node[below] {$A_3$} -- (0,0);
     \draw [MidnightBlue,very thick] (1.5,-1.5)node[below] {$B_1$} -- (2,0);
     \draw [MidnightBlue,very thick] (2.5,-1.5) node[below] {$B_2$}-- (2,0);
     \draw[very thick] (0,0) to (0,0.5);
     \draw[very thick] (2,0.5) to (0,0.5);
     \draw[very thick] (2,0) to (2,0.5);
     \draw[very thick] (2,0.5) to (3,0.5);
     
     \draw[very thick] (4,0.5) to (5,0.5);
     \draw[very thick] (5,0) to (5,0.5);
     \draw[very thick] (4,-1.5)node[below] {$M_1$} to (5,0);
     \draw[very thick] (4.66,-1.5)node[below] {$M_2$} to (5,0);
     \draw[very thick] (5.33,-1.5)node[below] {$M_3$} to (5,0);
     \draw[very thick] (6,-1.5)node[below] {$M_4$} to (5,0);
     \draw [fill,color=Maroon] (0,0) circle [radius=0.06];
     \draw [fill,color=MidnightBlue] (2,0) circle [radius=0.06];
     \draw [fill,color=black] (5,0) circle [radius=0.06];
     \draw [fill,color=black] (2,0.5) circle [radius=0.06];
     \node at (3.5,0) {\Large......};
    \end{tikzpicture}
    \caption{The geometry that dominates the partition function $Z_{\alpha^{m-1}}$ of the $m$-partite entropy $S^{(m)}(A:B:~...~:M)$, where pages $A_1,A_2,A_3$ belong to the subsystem $A$, etc. The minimal $m$-partite surface contains $m-1$ interfaces $\mathcal{S}^{(i)}$. }
    \label{fign-mentropy}
\end{figure}
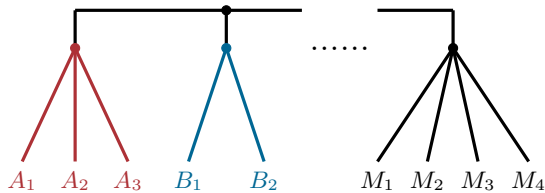

The preceding calculation naturally extends to the $m$-partite entropy for $n$ pages. While the on-shell booklet wormhole computes $Z_1$, the $Z_{\alpha^{m-1}}$ is evaluated through a modified geometry, as \cref{fign-mentropy} illustrates. The resulting $m$-partite holographic entropies are then \be S^{(m)}=(m-1)\frac{A_h}{4G},\ee which match the entropies of the GHZ states.

In fact, the introduction of $\epsilon$ is merely auxiliary; the real novelty originates from the ambiguous topology inherent in the junction of four or more partitions, a feature absent in conventional geometries.  Although the booklet wormhole is a single on-shell solution, it inherently incorporates distinct topological configurations. Consequently, the gravitational path integral for entropy calculations must sum over these topologies. This is a \textbf{non-perturbative quantum effect}: diverse paths must be considered even in the large-$N$/small-$G$ limit. This non-classical feature enables the evasion of the holographic entropy inequality \eqref{ineq} derived from classical geometries.

\section{Discussion}
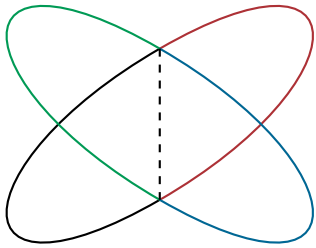
\begin{figure}
    \centering
    \begin{tikzpicture}[scale=0.5]
     \useasboundingbox (-3,-3) rectangle (3,3);
        \draw[thick, Maroon] (0,2) to[out=30, in=30,looseness=4]  (0,-2);
        \draw[thick, black] (0,2) to[out=210, in=210,looseness=4]  (0,-2);
        \draw[thick, MidnightBlue] (0,2) to[out=-30, in=-30,looseness=4]  (0,-2);
        \draw[thick, ForestGreen] (0,2) to[out=-210, in=-210,looseness=4]  (0,-2);
        \draw[thick,dashed] (0,2) to (0,-2);
    \end{tikzpicture}
    \caption{The sketch illustrates a constant-time slice of a booklet geometry, which also exhibits a non-perturbative quantum effect in the computation of entanglement entropies. Solid lines represent the CFT boundaries, and the dashed line denotes the multipartite junction in the bulk.}
    \label{anotherbooklet}
\end{figure}

The GHZ state/booklet wormhole duality is the first example that violates the conventional holographic entropy inequality \eqref{ineq} through a non-perturbative quantum effect. This violation occurs when the minimal entropy surface coincides with a multipartite junction. While prior works have employed cut-and-paste operations on AdS black holes \cite{24CouSta,24OuEqSt,24MicEnt}, the resulting minimal surfaces typically coincide with horizons in the absence of junctions; also, the junctions are not multi-way. Therefore, the topology of the minimal surface is robust. In contrast, positioning a multipartite junction at the minimal surface introduces an ambiguity in the minimal surface's topology, as we demonstrated in \cref{nonclassical}. Building on this idea, we can glue multiple RT surfaces together to reproduce a similar effect. For instance, consider constant-time slices of $n$ vacuum AdS spacetimes. We cut each slice along one geodesic and glue the resulting cuts together. The corresponding CFT points where these geodesics terminate are likewise identified. The resulting geometry is illustrated in \cref{anotherbooklet}. We will leave the exploration of this type of geometry for future study.

Multi-boundary wormholes with smooth manifold geometry have been studied in prior works \cite{14MulWor,15MulWor}. These geometries typically break $S_n$ permutation symmetry, with their dual CFTs having complicated wavefunctions and non-positive tripartite information. In principle, we can construct a continuous family of states interpolating between some state $I_3\leq 0$ and the GHZ state, whose dual geometries correspondingly vary from smooth multi-boundary wormholes to the booklet wormholes. If we imagine putting more and more GHZ-type particles into a smooth multi-boundary wormhole, one expect the booklet junction tp emerge by the time $I_3$ reaches zero. How this transition happens is an interesting open question.

While the Bell state is the most popular state in quantum information, the GHZ state exhibits novel features beyond it. For instance, unlike the Bell state which reveals non-locality statistically, Ref.~\cite{90GHZ} demonstrated that the GHZ state can exhibit non-locality in a single-shot experiment. How non-locality\textemdash and more fundamentally, quantum contextuality \cite{67KSTheo,05SpeCon}\textemdash{}manifests in holography remains an intriguing open question.

The ``ER$=$EPR'' conjecture \cite{13ER=EPR} proposes that quantum entanglement between two black holes is geometrically dual to an Einstein-Rosen bridge connecting their interiors. The booklet wormholes are also connected, but by a geometry different from that of the two-sided black holes.
On the quantum side, the GHZ states exhibit multi-partite entanglement, yet pairwise subsystems only show classical correlations. One might therefore expect that a GHZ wormhole may not connect observers falling into any pair of the black holes \cite{16GHZbla}. This work suggests the opposite: observers accessing two of the black holes \textbf{can} detect entanglement if they enter the horizons. We will discuss what observers falling into a booklet wormhole see in a subsequent paper \cite{26NoLoJu}, and explore how holographic teleportation \cite{17TraWor,17DouTra} is applied to booklet wormholes. 
This new model suggests that spacetime behind horizons geometrically embodies quantum entanglement (and hopefully, the hidden variables in contextuality). We hope this holographic model of many-body entanglement can provide a deeper understanding of ``spacetime from entanglement'' \cite{10SpaEnt}.
\\
\vspace{0.3cm}
\subsection*{Acknowledgments}
This work is supported by the National Natural Science Foundation of China grant No. 12375041. 
%%%%%%%%%%%%%%%%%%%%%%%%%%%%%%%%%%%%%%%%%%%%%%%%

\bibliography{ref}

\end{document}